\input harvmac
\newcount\figno
\figno=0
\def\fig#1#2#3{
\par\begingroup\parindent=0pt\leftskip=1cm\rightskip=1cm
\parindent=0pt
\baselineskip=11pt
\global\advance\figno by 1
\midinsert
\epsfxsize=#3
\centerline{\epsfbox{#2}}
\vskip 12pt
{\bf Fig. \the\figno:} #1\par
\endinsert\endgroup\par
}
\def\figlabel#1{\xdef#1{\the\figno}}
\def\encadremath#1{\vbox{\hrule\hbox{\vrule\kern8pt\vbox{\kern8pt
\hbox{$\displaystyle #1$}\kern8pt}
\kern8pt\vrule}\hrule}}

\overfullrule=0pt

\Title{TIFR-TH/98-50}
{\vbox{\centerline{Brane Waves, Yang-Mills theories and Causality}}}
\smallskip
\centerline{Sumit R. Das\foot{E-mail: das@theory.tifr.res.in}}
\smallskip
\centerline{\it Tata Institute of Fundamental Research}
\centerline{\it Homi Bhabha Road, Bombay 400 005, INDIA}
\bigskip

\medskip

\noindent

We provide evidence for the validity of AdS/CFT correspondence in the
Coulomb branch by comparing the Yang-Mills effective action with the
potential between waves on two separated test 3-branes in the presence
of a large number of other 3-branes.  For constant gauge fields
excited on the branes, this requires that the supergravity potential
in a $AdS_5 \times S^5$ background is the {\it same} as that in flat
space, despite the fact that both propagators and couplings of some
relevant supergravity modes are different.  We show that this is
indeed true, due to a subtle cancellation.  With time-dependent gauge
fields on the test branes, the potential is sensitive to retardation
effects of causal propagation in the bulk.  We argue that this is
reflected in higher derivative (acceleration) terms in the Yang-Mills
effective action.  We show that for two 3-branes separated in flat
space the structure of lowest order acceleration terms is in agreement
with supergravity expectations.

\Date{December, 1998}

\def\adsf{{AdS_5 \times S^5}}
\def\roR{({r \over R})}
\def\Ror{({R \over r})}
\def\vm{{\vec m}}
\def\half{{1\over 2}}
\def\vr{{\vec z}}
\def\tJ{{\tilde J}}
\def\cO{{\cal O}}
\def\cL{{\cal L}}
\def\cP{{\cal P}}
\def\tF{{\tilde F}}
\def\g{\delta t}

\newsec{Introduction.}

The duality between large-$N$ strongly coupled ${\cal N} = 4$ Yang-Mills theory
in $3+1$ dimensions and supergravity in a $AdS_5 \times S^5$
background \ref\malda{J. Maldacena, hep-th/9711200}
\ref\gkpw{S.S. Gubser, I.R. Klebanov and A.M. Polyakov, 
Phys. Lett. B428 (1998) 105, hep-th/9802109; E. Witten,
hep-th/9802150.} has served as an appropriate setting for discussing
holographic nature of theories of gravity. Traditionally this duality
is conjectured for the Higgs branch of the Yang-Mills theory,
corresponding to a large number of coincident three branes,
though the possibility that this could be valid in the Coulomb
branch - corresponding to branes separated from each other - 
was suggested already in \malda. More recently,
it has been argued by Douglas and Taylor \ref\dtay{M. Douglas and
W. Taylor, hep-th/9807225} that the duality indeed holds for the Coulomb
branch with brane positions identified with
Higgs vacuum expectation values.

If the correspondence is valid for the Coulomb branch, one would
be able to get new insight into the description of {\it local}
physics in the bulk in terms of Yang-Mills theory and thus
eventually understand black hole complementarity. Attempts
to understand motion of brane probes have been made in
\ref\holo{T. Banks, M. Douglas, G. Horowitz and E. Martinec,
hep-th9808016; V. Balasubramanium, P. Kraus, A. Lawrence and
S. Trivedi, hep-th/9808017.}. Some evidence for AdS/CFT
correspondence in the Coulomb branch has appeared in
\ref\trivnew{A. Bilal and C. Chu, hep-th/9810195; 
P. Kraus, F. Larsen and S. Trivedi, hep-th/9811120.} and a different
point of view is discussed in \ref\berkooz{M. Berkooz, hep-th/9807230.}

Consider for example a Higgs vev in the ${\cal N} = 4$ theory of the form
\eqn\one{\pmatrix{z_1^i & 0 & 0 \cr 0 & z_2^i & 0 \cr
0 & 0 & {\bf 0}_{(N-2)\times (N-2)}}}
The proposal is to identify $z_1^i$ and $z_2^i$ with
the transverse positions of a pair of three branes in the presence
of $(N-2)$ other branes - with all the branes parallel to each 
other. 

Now consider exciting this pair by turning on gauge fields $F_1$ and
$F_2$ respectively. At strong 't Hooft coupling, the low energy
effective action for these fields should then give the interaction
energy between the branes. Non-renormalization theorems
\ref\dine{M. Dine and N. Seiberg, hep-th/9705057;
S. Paban, S. Sethi and M. Stern, Nucl. Phys. B534 (1998) 137,
hepth/9805018 and JHEP 06 (1998) 012, hep-th/9806028.} may be then
used to calculate this energy by performing a one-loop computation for
special brane waves like those made of constant gauge fields.
The general one-loop answer for the $O(F^4)$ term 
is given by \ref\gates{S.J. Gates,
M.T. Grisaru, M. Rocek and W. Seigel,
``Superspace'', (Benjamin Cummings, 1987)}
\ref\peri{V. Periwal and R. von Unge, hep-th/9801121.} \dtay\
\eqn\two{\eqalign{ \int \prod_i d^4p_i & [ F^\mu_\nu (p_1)
F^\nu_\rho (p_2) F^\rho_\kappa (p_3) F^\kappa_\mu (p_4) - {1\over 4}
F^\mu_\nu (p_1) F^\nu_\mu (p_2) F^\rho_\kappa(p_3) F^\kappa_\rho (p_4)]\cr
&\delta^4 (\sum_{i=1}^4 p_i)~[G (p_1,p_2,p_3,p_4)
+ {\rm permutations}]}}
with $F = F_1 - F_2$ and
\eqn\three{G(p_i) = \int d^4 k [(\rho^2 + k^2)
(\rho^2 + (k-p_1)^2)(\rho^2 + (k-p_1-p_2)^2)(\rho^2
+ (k+p_4)^2)]^{-1}}
and
\eqn\four{ \rho^2 = \sum_{i=1}^6 (z^i_1 - z^i_2)^2}
We can expand $G(p_i)$ around $p_i = 0$ and obtain the position
space effective action in a derivative expansion. The first term is
\eqn\kthree{ {1\over \rho^4}\int d^4 y~[ F^\mu_\nu
F^\nu_\rho  F^\rho_\kappa  F^\kappa_\mu  - {1\over 4}
F^\mu_\nu  F^\nu_\mu  F^\rho_\kappa F^\kappa_\rho]}
This one loop contribution is exact for $N=2$ \dine. It is also exact
for $SU(N)$ in the Coulomb branch where all the higgs have nonzero
expectation values in which case the expression \two\ should include
a sum over all $U(1)$ factors
\ref\lowe{D. Lowe, hep-th/9810075; D. Lowe and R. von Unge,
hep-th/9811017}
. In \dtay\ it has been argued that
the nonrenormalization theorems should still hold for our case, 
where $SU(N) \rightarrow
SU(N-2) \times [U(1)]^2$, by
considering \one\ as a limit from the Coulomb branch - though there
is no proof as yet. This situation
could be therefore compared to a supergravity calculation. In this
limit, \two\ leads to an effective potential which has the
following terms
\eqn\five{{1\over \rho^4}[O^\phi_1 O^\phi_2 + O^\chi_1 O^\chi_2 + 
2 T^{\mu\nu}_1 T_{2~\mu\nu} + 2 O^{\mu\nu}_1 O_{2~\mu\nu}|_4]}
where for each $i = 1,2$ 
\eqn\zfive{\eqalign{& O^\phi_i = {1\over 4}(F_i)^{\mu\nu}(F_i)_{\mu\nu}\cr
& O^\chi_i = {1\over 4}(F_i)^{\mu\nu}(\tF_i)_{\mu\nu}\cr
& T^{\mu\nu}_i = {1\over 2}[(F_i)^\mu_\alpha (F_i)^{\alpha\nu}
- {1\over 2} \eta^{\mu\nu} (F_i)^{\alpha\beta}(F_i)_{\alpha\beta}]\cr
& O^{\mu\nu}_i = {1\over 2}[(F_i)^{\mu\nu} + (F_i)^{\mu\alpha}
(F_i)_{\alpha\beta}(F_i)^{\beta\nu} - {1\over 4} (F_i)^{\mu\nu}
(F_i)^{\alpha\beta}(F_i)_{\alpha\beta}]}}
In \five\ the subscript $|_4$ in the last term means that we retain
terms containing four factors of the gauge field in the product.

Let us first consider the case where the gauge fields on the branes
are constants. Then \kthree\ is the only contribution to the
effective action upto $O(F^4)$.

On the supergravity side, the force between the pair is due to the
exchange of supergravity modes. With only the gauge field excited
these modes are the dilaton, axion, longitudinally polarized graviton
and the longitudinally polarized 2-form fields.
When $N = 2$ this is propagation 
in flat space.
In this case it is straightforward to understand the terms in
\five. The overall factor $1/\rho^4$ comes from the static
massless propagator
in the six transverse dimensions, which appears because the
supergravity modes which couple to constant gauge fields on the
brane have zero momentum along the brane directions.
The first term is due to the exchange of a dilaton, which couples to
${\rm Tr} F^2$ on each of the branes \ref\igora{I.R. Klebanov,
Nucl. Phys. B496 (1997) 231, hep-th/9702076; S.S. Gubser,
I.R. Klebanov and A.A. Tseytlin, Nucl. Phys. B499 (1997)
217, hep-th/9703040.}. The second term comes from axion exchanges
which couples to ${\rm Tr}(F\tF)$.
The third term comes from the exchange of
a longitudinally polarized graviton which couples to the energy
momentum tensor $T_{\mu\nu}$ on the worldvolumes \igora. The last term
comes from the exchange of a 2-form field, whose couplings have
been obtained in \ref\dtrivedi{S.R. Das and S.P. Trivedi,
hep-th/9804149}. (Here the 2-form couples to $F_{\mu\nu}$ on one of
the branes and to a cubic in the fields on the other one). Moreover,
the relative coefficients between the various operators in \five\
are exactly what is expected from the couplings and propagators. 
The fact that the supergravity answer agrees with the gauge
theory truncation of the open string theory on the brane is well
known in related contexts \ref\dkps{M. Douglas, D. Kabat, P. Pouloit
and S. Shenker, Nucl. Phys. B485 (1997) 85, hep-th/9608024;
M. Douglas, J. Polchinski and A. Strominger, JHEP 12 (1997) 003,
hep-th/9703031.}. 

For large $N$ and in the scaling limit, however, the pair of branes
are situated in the $AdS \times S^5$ produced by the $N-2$ other
branes and one has to use the couplings and propagators in this
space-time. It is puzzling how the same Yang-Mills
answer in \five\ could be reproduced by supergravity in a
nontrivial space-time. In particular, the flat space propagator
$1/\rho^4$ depends on the coordinate distance between the
branes and not on their individual locations - a feature which is
not {\it a priori} expected in $AdS_5 \times S^5$.

Remarkably, as was shown in \dtay, the $AdS_5 \times S^5$ propagator
for fields which obey the massless Klein-Gordon equation is identical
to the flat space propagator when restricted to the zero brane
momentum sector. Examples of such fields are the dilaton and the
longitudinally polarized graviton. Moreover, as is clear from the
analysis of \dtrivedi, the couplings of these fields to the
individual branes are the same as that in flat space. Thus the
first three terms in \five\ indeed follow from dilaton and graviton
exchanges in $\adsf$.

In \dtay, it was claimed that the last term  of \five\ can be also
explained by 2-form exchange in $\adsf$.  However, because of the
presence of a nonzero 5-form field strength in the $\adsf$ background
, the NS-NS and the R-R 2-forms mix with each other through a
Chern-Simons term in IIB supergravity \ref\van{H.J. Kim, L. Romans
and P. van Nieuwenhuizen, Phys. Rev. D32 (1985) 389.},
 leading to two
independent branches and these branches behave as {\it massive}
fields. Various other supergravity modes mix with each other in a
similar fashion.  This phenomenon is crucial in a supression of the
classical s-wave absorption cross-section of the 2-form field by three
branes \ref\rajaraman{A. Rajaraman, hep-th/9803082.}. Consequently, as
will be shown below, this leads to rather different propagators which
reflect the mixings and also depend on individual brane locations.
Moreover, as shown in \dtrivedi, the coupling of
the 2-form fields to the brane are {\it different} in $\adsf$ and flat
space. It would be rather miraculous if inspite of such differences,
the supergravity calculation is able to reproduce \five.

In this paper we show that this miracle indeed happens. The difference
in couplings and the propagators conspire to reproduce the exact form
of the two-form mediated potential expected from Yang-Mills theory.
We conjecture that this mechanism is quite general and would be
manifest in the interaction between other brane waves which involve
exchange of other supergravity modes displaying a similar
mixing. Our results provide strong evidence for the validity of
Maldacena conjecture in the Coulomb branch.

Finally we address the question of causality in the bulk and its
manifestation in the Yang-Mills effective action.
In supergravity, the interaction between test branes occurs through
{\it retarded} potentials arising from causal propagation of supergravity
modes. For constant gauge fields on the branes, retardation effects
are invisible and static propagators in transverse space are relevant.
However, for nonconstant waves, causality manifests itself by producing
an interaction energy which is bi-local on the brane
\foot{This point has been emphasized to me by S. Mathur}. 
From the point of view of the AdS/CFT correspondence, it may appear
puzzling how the boundary Yang-Mills
theory ``knows'' about causality in the bulk. In particular when the
two test branes are separated only in the radial direction, the two
locations map into the {\it same} point on the boundary and causality
in the boundary theory does not impose any restriction.
In fact, the Yang-Mills effective action is usually written as a sum
of {\it local} terms. 

We will argue that bulk causality is reflected in the Yang-Mills
theory in terms involving derivative of the fields.  Supergravity then
predicts a specific structure of these terms. We show this explicitly
for the lowest order acceleration terms involving gauge fields in the
case of two test branes in flat space by comparing the result with the
effective action of $SU(2)$ Yang-Mills theory.  Fortunately this term
is not renormalized, thus a comparison with supergravity is
allowed. We expect that this will continue to hold in $AdS_5 \times
S^5$, which we will discuss in a future publication
\ref\dmn{S.R. Das and S.D. Mathur, {\it to appear}}. In general such
considerations may lead to a supergravity understanding of the
``acceleration'' terms in the Yang-Mills effective action.

\newsec{Propagators at zero brane momentum}

Consider the following metric in $\adsf$
\eqn\six{ ds^2 = \roR^2  [dy \cdot dy] + \Ror^2
[dr^2 + r^2 \sum_{i=1}^5 f_i(\theta_i) (d\theta_i)^2]}
We will use the following conventions. The ten dimensional
coordinates will be denoted by $y^a, a=0,\cdots 9$. Out of
these we continue to denote the brane worldvolume directions
by $y^\mu, \mu = 0,\cdots 3$.
The remaining six transverse coordinates
$y^5 \cdots y^9$ will be relabelled as $z^i, i = 1,\cdots 6$.
$r = {\sqrt{\sum_{i=1}^6 (z^i)^2}}$ is the
radial coordinate in the transverse space and $\theta_i$ are
angles on the $S^5$. $(r, \theta_i)$ are related to the cartesian
coordinates $z^i$ in the transverse space by the standard
transformations and the metric coefficients $f_i (\theta_i)$ are
determined from these transformations. 

In the following we will set $R = 1$ without loss of generality,
and restore them using dimensional analysis when required.

The action for a minimally coupled massless scalar in this background
may be easily seen to be
\eqn\seven{S_\phi = {1\over 2} \int dt~d^3x~d^6z [{1\over r^4}
 (\partial_{x^\mu} \phi)^2 + (\partial_{z^i} \phi)^2]} 
Thus when the fields do not depend on the brane worldvolume coordinates,
the action is in fact identical to that of a massless scalar field
in flat space. This means that the propagator with zero worldvolume
momentum is the flat space propagator and given by
\eqn\eight{G_\phi (z_1, z_2) = {1\over 4\pi^3 |z_1 - z_2|^4}}
The dilaton and the longitudinally polarized graviton 
\foot{Longitudinal polarization means that the tensor indices of
the fields are along the 3-brane worldvolume.} behave like
massless minimally coupled scalars from the point of view of the
six dimensional transverse space and therefore has propagators
given by \eight. This is the result used in \dtay.

The longitudinally polarized 2-form field is also a scalar from
the point of view of the transverse space, but it is not a
minimally coupled scalar. More significantly, there are two such
2-form fields, the NS-NS field which we denote by $b_{ab}$ and
the R-R field which we will denote by $a_{ab}$. These two fields
are coupled with each other through the background five form
field strength. It is conveninent to combine these two fields into
a single complex field
\eqn\nine{B_{ab} = b_{ab} + i a_{ab}}
the relevant part of the supergravity action \foot{In this paper
we do not consider fluctuations of the five form field strength,
so that usual problems of writing an action for a self dual
five form gauge field are not relevant.}
is given by
\eqn\ten{S_B = {1\over 12}\int d^{10} x~{\sqrt g}[ H_{abc}^* H^{abc}
+ i F^{abcde}(H_{abc}B_{de} - H^*_{abc} B_{de} - (c.c.))]}
where
\eqn\eleven{H_{abc} = \partial_a B_{bc} +\partial_b B_{ca}
+ \partial_c B_{ab}}
The last term is a Chern-Simons term which couples the two types
of fields. This leads to the well known equations of motion
\eqn\twelve{ {1\over {\sqrt g}}\partial_c({\sqrt g} H^{cab})
= - {2i \over 3} F^{abcde}H_{cde}}

In the $\adsf$ background the five form field has a value
\eqn\thirteen{F^{12r03} = {1 \over r^3}}
the other nonzero components being determined by antisymmetry
and self-duality in the usual fashion. 

We are interested in the longitudinal components of the 2-form
field, so that in $B_{ab}$ the indices $(a,b)$ take values
$a,b = 0,\cdots 3$. The equations \eleven\
and \twelve\ then show that a given component $B_{\mu\nu}$ mixes
only with its dual ${1\over 2}\epsilon_{\mu\nu\alpha\beta}B^{\alpha
\beta}$. It is therefore convenient to define three pairs of 
complex fields $(\phi^A_1,\phi^A_2), A = 1\cdots 3$
denoting the electric and magnetic parts of $B_{\mu\nu}$
\eqn\fourteen{\phi^A_1 = {1\over 2}\epsilon^{ABC}B_{BC}
~~~~~~~~~~\phi^A_2 = B_{0A}}
We fix a gauge in which the fields are independent of the coordinates
$y^\mu, \mu = 0,\cdots 3$. We also introduce a coordinate
\eqn\fifteen{ x = {\rm log~}r}
The action \ten\ for fields which depend only on the transverse
coordinates $z^i$ in the background given by \six\ and
\thirteen\ then becomes
\eqn\sixteen{\eqalign{S_B = {1\over 2}\int dx~[d\Omega_5]\sum_{A=1}^3
[&\partial_x \phi^{A~*}_1 \partial_x \phi^A_1
+\sum_{i=1}^5{1\over  f_i}\partial_i \phi^{A~*}_1
\partial_i \phi^A_1\cr
&-\partial_x \phi^{A~*}_2 \partial_x \phi^A_2
-\sum_{i=1}^5{1\over  f_i}\partial_i \phi^{A~*}_2
\partial_i \phi^A_2\cr
&+4i(\phi^{A~*}_1 \partial_x \phi^A_2 +
\phi^{A~*}_2 \partial_x \phi^A_1)]}}
Here the measure on $S^5$ is given by
\eqn\seventeen{d\Omega_5 = (\prod_{i=1}^5 d\theta_i)~h(\theta_i)
~~~~~~~h(\theta_i) = (f_i)^{1\over 2}}
The negative signs in the kinetic terms of $\phi^A_2$ come
from lowering a timelike index.

Clearly the action \sixteen\ is {\it not}
the same as that in flat space, unlike the minimally coupled
scalar discussed above. This, together with the mixing between
$\phi^A_1$ and $\phi^A_2$ makes the propagator nontrivial.
Furthermore the different pairs $\phi^A_n$ are independent 
of each other and may be treated separately.

The propagators for these fields may be obtained by performing
a standard mode decomposition to diagonalize the action. The
details are given in Appendix B. The final result for the propagator
is, after restoring factors of $R$
\eqn\twosixa{N^{AB}(\vr_1,\vr_2) = {\delta^{AB}\over 
8 \pi^3 R^4 |\vr_1 - \vr_2|^4}
\pmatrix{(r_1^4 + r_2^4) & -
i(r_1^4 - r_2^4) \cr
-i(r_1^4 - r_2^4) & 
-(r_1^4 + r_2^4)}}
The propagator may be, of course, 
expressed in terms of geodesic distances. However that will not
be necessary for our present purposes.

It may be easily checked that the propagators for the 2-form fields in
{\it flat} space is 
\eqn\twoseven{N^{AB}_{{\rm flat}} = {\delta^{AB}\over 
4 \pi^3 |\vr_1 - \vr_2|^4}
\pmatrix{1 & 0 \cr 0 & -1}}
in sharp contrast with \twosixa. The relative factor of $2$ in the
overall normalizations in \twosixa\ and \twoseven\ will be crucial
in what follows.

Finally let us consider current couplings in the supergravity
theory with currents $J(x,\theta_i)$ which depend only on the
transverse directions
\eqn\twoeight{{1\over 4}\int dx d\Omega_5
\sum_{A,n}[(J^A)^*_n(x,\theta_i)
\phi^A_n (x,\theta_i) + J^A_n(x,\theta_i)(\phi^A)^*_n 
(x,\theta_i)]}
Integrating out the fields one gets the current-current coupling
\eqn\twonine{{1\over 4} \int dx d\Omega_5 \int dx' d\Omega'_5
[(J^A)^*_n (x,\theta)N^{AB}_{nm} (x,\theta;x', \theta') 
J^B_m(x',\theta')]}
Note that reality requires
\eqn\thirty{(N^{AB})^*_{mn} (x,\theta;x',\theta') = N^{BA}_{nm}
(x',\theta';x,\theta)}
which is satisfied by our propagator \twosixa.

\newsec{Couplings in the Dirac-Born-Infeld-Wess-Zumino action}

The couplings of the relevant supergravity modes to a single brane in
$\adsf$ may be obtained from the Dirac-Born-Infeld-Wess-Zumino
(DBI-WZ) action, and have been studied in \dtrivedi. The action for
a D3-brane in a general background of dilaton, graviton and rank-2
fields is given by \ref\ceder{M. Cederwall, A. von Gussich, 
B. E. W. Nilsson and 
A. Westberg, Nucl. Phys. B490 (1997) 163, hep-th/9610148; 
E. Bergshoeff and
P.K. Townsend, Nucl. Phys. B490 (1997) 145.}
\eqn\fourone{S=-\int d^4 \xi {\sqrt { -det(G_{\mu\nu} +
{\cal F}_{\mu\nu} )} } + \int ({\hat
C_{(4)}} + {\cal F} \wedge {\hat A} + {\hat C_{(0)}} 
{\cal F} \wedge {\cal F} )}
\ref\aps{M. Aganagic, C. Popescu and J. H. Schwarz, , hep-th/9612080; 
M. Aganagic, J. Park, C. Popescu and J. Schwarz, hep-th/9702133.}
The two terms above correspond to the DBI action and the
WZ term respectively. $G_{\mu\nu}$ 
refers to the induced world-volume metric,
obtained as the pull-back of the spacetime metric. Similarly,
\eqn\fourtwo{{\cal F}_{\mu\nu}= F_{\mu\nu} - {\hat  B_{\mu\nu}}}
where $F_{\mu\nu}$ stands for the gauge field on the D3-brane and ${\hat
B_{\mu\nu}}$ is the pullback of the NS-NS two form potential. In the W-Z
term ${\hat C_{(4)}}$,  ${\hat A }$ and ${\hat C_{(0)}}$ refer to the
pullback of the R-R four form,  
two form and zero form fields  respectively. The DBI-WZ action may
be viewed as the effective action of the Yang-Mills theory
with $SU(N) \rightarrow SU(N-1) \times U(1)$,
with derivatives on gauge fields ignored.
The diagonal
Higgs which breaks the symmetry interpreted as the position of a
3-brane probe. Conformal transformations of the Higgs fields in
the Yang-Mills description are ``metamorphosed'' into those of
the transverse coordinates in $\adsf$ due to modifications of
Ward identities in the gauge fixed theory \ref\jky{A. Jevicki,
Y. Kazama and T. Yoneya, Phys. Rev. Lett. 81 (1998) 5072,
hep-th/9808039; A. Jevicki, Y. Kazama and T. Yoneya, hep-th/9810146.}.

We fix a static gauge, setting the four worldvolume parameters to
be equal to the coordinates $y^\mu$ in the metric and also 
fix the kappa symmetry following \aps\ by setting half of the
fermionic fields in the brane action to zero. The couplings of the
various supergravity modes may be then obtained by performing an
expansion around background values (given by the $\adsf$ solution
and the five form background field strength) and then expanding the
determinant to the required order.

We will consider the case when only the bosonic gauge fields are
excited on the brane. 
Then the operator on the worldvolume which couples
to the dilaton obtained by the above procedure is
\eqn\fourthree{\cO_\phi = -{1\over 4} F^{\mu\nu}F_{\mu\nu}}
where indices of worldvolume fields are raised and lowered using the
flat metric.
The same operator couples to a brane in flat space \igora. Similarly
the operator coupling to the longitudinal components of the metric
is also of the same form as in flat space
\eqn\fourfour{(\cO_g)^{\mu\nu} = {1\over 2}[F^{\mu\alpha}F_\alpha^\nu
- {1\over 4} \eta_{\mu\nu}(F^{\mu\nu}F_{\nu\mu})]}
In contrast the operator coupling to the antisymmetric tensor fields
$a_{\mu\nu}$ and $b_{\mu\nu}$ differ in important detail from their
form in flat space. The operator for the NS-NS form comes from the DBI
term and is given by \dtrivedi\
\eqn\fourfive{(\cO_b)^{\nu\mu} = -\half[F^{\nu\mu}
+ {1\over r^4} G^{\nu\mu}]}
where
\eqn\fourfivea{ G^{\nu\mu} = 
[F^\nu_\rho F^\rho_\kappa F^{\kappa \mu}
- {1\over 4} (F^{\rho\kappa}F_{\rho\kappa})F^{\nu\mu}]}
while that for the R-R field comes from the WZ term
\eqn\foursix{(\cO_a)^{\nu\mu} = {1\over 4}\epsilon^{\nu\mu\rho\kappa}
F_{\rho\kappa}}
In \fourfive\ $r$ is the location of the brane in question.

If the brane was located in flat space one would simply have
\eqn\fourfiveb{(\cO_b)^{\nu\mu} |_{\rm flat} = -\half[F^{\nu\mu}
+ G^{\nu\mu}]}
The factor of $1/r^4$ in front of $G^{\nu\mu}$ is now absent.

In \dtrivedi\ it was shown that the operator \fourfive, modified
by the prescription of \ref\tsey1{A. Tseytlin,
Nucl. Phys. B501 (1997) 41, hep-th/9701125.}, for the
nonabelian analog, represents
the 2-form field in the dual description in terms of a Yang-Mills
theory. Note that for this to hold the supergravity modes 
to which they couple have to be {\it on shell}. This played
a crucial role in cancellation of dimension four
fermionic operators, which would have jeopardazied the
AdS/CFT connection.

For our purposes such an interpretation is not necessary -
we will simply consider these operators for what they stand :
coupling of individual branes to supergravity modes. By the
same token we remain {\it off-shell}.

The presence of the factor of $1/r^4$ (which is actually
$(R/r)^4$ once the $R$ is restored) in front of the dimension
six term is related to the relationship between the infrared
cutoff in $AdS$ space and the ultraviolet cutoff in the dual
gauge theory - a fact that is crucial for holography
\malda,\ref\suswit{L. Susskind and E. Witten, hep-th/9805114;
A. Peet and J. Polchinski, hep-th/9809022},\holo. 
When the dual theory is considered to live on
the boundary at large $r$ this term may be thought of
providing the ultraviolet cutoff necessary to write down
a higher dimension operator in the gauge theory.
The presence of this factor of $1/r^4$ in \fourfive\ will turn
out to be crucial in what follows.

Note the asymmetry between the NS-NS and R-R fields in the couplings.
The 3-brane is of course self-dual. In the dual formulation, the
NS-NS fields are interchanged with the R-R fields and the field
strength is replaced by its dual as well \ref\tseytlin{A. Tseytlin,
Nucl. Phys. B469 (1996) 51, hep-th/9602064.},\aps.

Because of the presence of the $\epsilon_{\mu\nu\rho\kappa}$ in
\foursix, it is natural to rewrite the coupling
\eqn\fourseven{ \cL_I = b_{\mu\nu}(\cO_b)^{\nu\mu}
+ a_{\mu\nu}(\cO_a)^{\nu\mu}}
in terms of the fields $\phi^A_n$ introduced in the previous
sections. 
\eqn\foureight{\cL_I = \half\sum_{A,n} [\phi^{A~*}_n \cP^A_n
+ c.c.]}
where we have
\eqn\fournine{\eqalign{&\cP^A_1 = \half[ \half \epsilon^{ABC}(F_{BC} 
+ {1\over r^4}G_{BC}) + i F^{0A}]\cr  
&\cP^A_2 = \half[(F^{0A} + {1\over r^4}G^{0A}) + 
{i \over 2}\epsilon^{ABC}F_{BC}]}}

In the full ten dimensional theory, the interaction of
the 2-form field with a pair of branes located at $\vr = \vr_1$
and $\vr = \vr_2$ 
may be then written as
\eqn\fifty{\int d^4 y\int d^6 z\sum_{A,n}[\phi^{A~*}_n (y,z) \tJ^A_n
(y,z) + c.c.]}
where
\eqn\fiveone{\tJ^A_n (y,z) = [\delta^6 (z-z_1) + \delta^6 (z-z_2)]
\cP^A_n (y,z)}
We are interested in the situation where the brane waves are constant
along the brane, so that the operators $\cP$ are independent of
$y$. In that case the $y$ integration in \fifty\ projects out the
zero brane momentum part of the fields $\phi^A_n$ and one is left
with an expression of the form \twoeight. Since the measure in
\twoeight\ is $dx d\Omega_5$ while that in \fifty\ it is
$dr~d\Omega_5~r^5$ one has
\eqn\fivetwo{ J^A_n (x,\theta) = r^6 \tJ^A_n (x,\theta)}

\newsec{Interaction Energy of constant field brane waves}

We can now use the formulae in Section 2. to derive the interaction
energy between brane waves due to 2-form exchange. This is the
connected piece in \twonine, where we substitute \fivetwo\ and
\fiveone. Note the additional factor of $r^6$ present in \fivetwo\
can be absorbed by changing the measure in \twonine\ to yield
\eqn\sixone{\int d^6z \int d^6 z' [\delta^6 (z-z_1) + \delta^6 (z-z_2)]
[\delta^6 (z'-z_1) + \delta^6 (z'-z_2)] \cP^{A~*}_m (z) N^{AB}_{mn} (z,z')
\cP^B_n(z')}
where $N^{AB}_{mn}$ is the zero momentum propagator which has been
calculated above. The interaction energy is given by the 
connected piece
\eqn\sixtwo{E = \cP^{A~*}_m(z_1)N^{AB}_{mn}(z_1,z_2)\cP^B_n(z_2)
+\cP^{A~*}_m(z_2)N^{AB}_{mn}(z_2,z_1)\cP^B_n(z_1)}
Evaluating \sixtwo\ using  \fournine\ and \twosixa\ is straightforward.
The final result is
\eqn\sixthree{E = {1\over 4\pi^3 \rho^4}[(F_1)^\mu_\nu
(F_2)^\nu_\rho(F_2)^\rho_\kappa(F_2)^\kappa_\mu -{1\over
4}(F_1)^\mu_\nu(F_2)^\nu_\mu(F_2)^\rho_\kappa(F_2)^\kappa_\rho + (1
\rightarrow 2)]} 
Using \fourfiveb\
and \twoseven\ it is easily seen that we get an {\it identical} result
for just two three branes located in flat space. The relative factor
of two in the overall normalizations of the flat space and AdS propagators
is crucial for this agreement.

\noindent
Two sets of important cancellations happened for each term over
the indices $(A,B)$

\item{1.} Terms quadratic in $F$'s, like 
$(F_1)^\mu_\nu(F_2)^\nu_\mu$, which could have been present because of
terms in $\cP^A_n$ linear in $F$, cancelled. If this did not happen,
there would be no correspondence with Yang-Mills. These
would be loop corrections to the kinetic energy terms,
which cannot be present in this $N=4$ theory.

\item{2.} Both the propagator and the couplings depend on the
individual brane locations $\vr_1$ and $\vr_2$. However these
translation-noninvariant terms conspire to cancel each other leaving
with an answer which depends only on $|\vr_1 - \vr_2| $.

\noindent
The structure in \sixthree\ is in {\it precise} agreement with the result
of Yang-Mills theory given in the last line \five. 

Since the couplings and zero momentum propagators for the dilaton
and the graviton are identical in $\adsf$ and flat space we would
trivially reproduce the first two lines of \five.

\newsec{Other brane waves}

Even for the simple brane waves considered above, i.e.
constant gauge fields, the agreement of Yang-Mills effective action in
Coulomb branch and the interaction between branes in supergravity
through single mode exchange depends on the non-trivial cancellation
demonstrated above. It is certainly worth understanding this
mechanism by studying other kinds of brane waves, e.g. excitations
of fermions or Higgs fields on the worldvolume.

Of particular interest are fermionic operators. These would couple to
the gravitons via their contribution to the energy momentum tensor and
to the two-form field via operators which have been derived in
\dtrivedi. It may be easily verified, using the 
nature of the 2-form propagators
derived above, that the fermionic operators do not have a net
contribution from 2-form exchange 
{\it both} in $\adsf$ as well as in flat space.
This again is due to cancellations, but now the contributions from
the diagonal and the off-diagonal parts of the propagators cancel
separately \foot{The same mechanism is responsible for the
{\it on shell} cancellation of dimension four fermionic operators
required for AdS/CFT correspondence to hold \dtrivedi.}.
The Yang-Mills contributions may be read off from
the results of \gates,\dine and \ref\rocek{B. de Wit, M.T. Grisaru and
M. Rocek, Phys. Lett. B374 (1996) 297, hep-th/9601115;
U. Lindstrom, F. Gonzalez-Rey, M. Rocek and R. von Unge,
hep-th/9607089;
F. Gonzalez-Rey and M. Rocek, Phys. Lett. B434 (1998) 303,
hep-th/9804010; F. Gonzalez-Rey, B. Kulik, I.Y. Park
and M. Rocek, hep-th/9810152.}.

When other brane waves are excited, various other supergravity 
modes will contribute to the exchange and {\it a priori}
their propagators would not be the same as in flat space.
For example, with the Higgs field excited, there is 
a coupling with the trace of the $S^5$ metric which mixes
with the rank-4 gauge field polarized along $S^5$ \van.
It would be interesting to see whether similar cancellations
hold in this case as well.

\newsec{Time dependent brane waves and causality in the bulk}

So far we have restricted our attention to interactions mediated by
supergravity modes with zero brane momentum. 
This restriction hides an important
piece of physics in the bulk : causality. The point is that the force
between any two objects is mediated by {\it retarded} propagators
reflecting causal propagation and not by instantaneous action. 
This does not have an obvious meaning in the Yang-Mills description.
The base space-time of Yang-Mills theory is
identified with the directions $y$ in the bulk, but there
is no analog of the radial distance $r$. Consider for example two
points which are separated in the $AdS_5$ space along the
radial direction. A physical signal takes a finite time to travel
between these points. However in the Yang-Mills description these
points are in fact the {\it same} point in the boundary space.
It seems rather mysterious as to how the
Yang-Mills theory encodes this finite time lag.

In the following we will argue that the supergravity prediction for
force between branes due to causal propagation of massless modes
leads to a precise prediction for the structure of higher derivative
operators in the effective action in the Yang-Mills theory.

Consider a general coupling to the test branes of the
form given by \fifty\ and \fiveone. For an arbitrary supergravity
field $\Phi_M$ this is given by
\eqn\afifty{\sum_M\int d^4 y\int d^6 z[\Phi^*_M (y,z) \tJ_M
(y,z) + c.c.]}
where
\eqn\fiveone{\tJ_M (y,z) = [\delta^6 (z-z_1) + \delta^6 (z-z_2)]
\cP_M (y,z)}
where indices $M$ label various supergravity fields.
The currents $\cP_M$ are made out of fields on the brane.
To illustrate the point, we will
consider currents $\cP_M$ which depend only on time. This, in fact,
highlights the issue since the coupling of the fields in \fifty\ are
to currents on the two branes which are at the same {\it spatial} position
on the brane. Then the interaction energy is given by the expression
\eqn\sone{E = \sum_{MN}\int dt \int dt'~\cP_M^* (t,z_1)~ \Delta^R_{MN} 
(t-t';z_1,z_2)~\cP_N (t',z_2)}
where $\Delta^R_{MN}$ denotes the retarded propagator, and
we have assumed time translation invariance. In the special
case considered in the previous sections, i.e. with 
time independent $\cP_M$ ,
the time integrals pass through the currents and convert the
retarded propagator into a static propagator in transverse space.
For general time dependence, this does not happen and one is left
with a bilocal expression for the interaction energy, given above.

The Yang-Mills effective action, however, is given as a sum of
various terms which are integrals of local densities on the brane
worldvolume. In our example this involves a single integral over
time since the fields are assumed to be constant in space.

Our proposal for comparing the supergravity and Yang-Mills expressions
is to expand the currents in \sone\ around the average time. Introducing
\eqn\stwo{t_0 = \half (t+t')~~~~~~~\g = t-t'}
we find from this Taylor expansion
\eqn\sthree{ \eqalign{E = \sum_{MN}[&\int dt_0 \cP^*_M(t_0,z_1) 
N_{MN} (z_1,z_2)
\cP_N (t_0,z_2)\cr
& + \half \int dt_0 (\partial_0 \cP^*_M(t_0))
(\partial_0 \cP_N(t_0))~\int [d(\g)]~(\g)^2~\Delta^R_{MN}(\g;z_1,z_2)
+ \cdots]}}
where the dots denote higher order terms in $\g$. 
The first term involves the static propagator $N_{MN}$ 
\eqn\sthreef{N_{MN} (z_1,z_2) = \int dt \Delta^R_{MN}(t;z_1,z_2)}
which we considered in the previous sections. However the currents
$\cP_M (t)$ are general functions of time. In the general case they
may be considered as general functions of the brane worldvolume
coordinates.  This explains how supergravity generates $F^4$ terms in
the effective action, as in \kthree, even when the fields are not
constant.

Since the currents $\cP_N$ are composite operators involving gauge
fields, the successive terms in \sthree\ should corrrespond to terms
in the Yang-Mills effective action which are higher order in a time
derivative expansion.  Moreover, as we will see shortly, the integral
over $\g$ converts the expansion in terms of the time lag into an
expansion in terms of the magntitude of the transverse distance. The
latter is, however, the magnitude of the Higgs expectation value and
hence the scale below which the low energy effective action is
valid. This is a direct manifestation of the IR-UV
correspondence. This has played a role in earlier discussions of bulk
causality \ref\kabat{D. Kabat and G. Lifschytz, hep-th/9806214.}.

Causality in the bulk therefore provides a specific structure for these
higher derivative terms for the strongly coupled Yang-Mills theory,
strong coupling being required for the 
validity of the supergravity approximation of  IIB string theory.
To check this proposal we need to find such operators which are
protected by non-renormalization theorems, so that we can perform
a weak coupling calculation in the gauge theory.

In fact the simplest test involves currents which are linear in
the gauge fields, which couples to the 2-form field in the bulk.
In terms of the notation introduced above we then have
\eqn\sfour{\eqalign{&\cP^A_1 (t,z) = \half \epsilon^{ABC} F_{BC} (t) 
+i F^{0A} (t) \cr
&\cP^A_2 (t,z) = F^{0A} (t) + i \half \epsilon^{ABC} F_{BC} (t)}}
The first term in \sthree\ then involves two powers of the gauge
field and no derivatives - these cancel as shown in section 4.
The second term in \sthree\ is of the form $(\partial F)^2$,
which is of ``weight'' four and hence protected by nonrenormalization
theorems of \dine. We will show soon that there is a net contribution
to these derivative terms.

\subsec{Branes in flat space and $SU(2)$}

The considerations of causality are equally relevant to the situation
with two separated three branes, with no other branes present.
In supergravity, these are then located in flat space and we
can evaluate the expressions easily. The Yang-Mills description
is then in terms of a $SU(2)$ gauge theory. 

As before we deal with the 2-form fields with polarizations along
the brane worldvolume. The fields are assumed to depend on time
and the transverse directions. We will work in 
a gauge $\partial^a B_{ab} = 0$.
The action for these modes in a
flat background can be easily worked out to be
\eqn\sfive{S = \half \sum_A \int dt \int d^6 z [ -|\partial_t \phi^A_1|^2
+ |\partial_z \phi^A_1|^2 - |\partial_z \phi^A_2|^2]}
where the fields $\phi^A_n$ have been defined in \fourteen. Note that
the action does not involve time derivatives of the ``electric''
components $\phi^A_2$, so that the propagator for this is essentially
the static propagator in transverse space. The contributions of
$\phi^A_1$ and $\phi^A_2$ to the first term in \sthree\ cancel,
essentially due to the negative sign of the kinetic term for
$\phi^A_2$ - as has been shown in the previous sections. In the
second term, only $\phi^A_1$ contributes and one is thus left
with a term
\eqn\ssix{\sum_A \int dt_0 (\partial_t \cP^A_1 (t_0))^*(\partial_t
\cP^B_1 (t_0)) 
\int d\g~(\g)^2~\Delta^{R,AB}_{11}(\g;z_1,z_2)}
where the retarded propagator for $\phi^A_1$ can be read off from
the action \sfive\
\eqn\sseven{ \Delta^{R,AB}_{11} (\g,z_1,z_2) = \delta^{AB}
\int {dp_0 \over 2\pi} \int {d^6 p \over (2\pi)^6}
{e^{-ip_0\g + i {\vec p}\cdot ({\vec z}_1 - {\vec z}_2)}
\over (p_0 + i\epsilon)^2 - ({\vec p})^2}}
The integral in \ssix\ may be easily seen to be
\eqn\seight{\int d\g~(\g)^2~\Delta^{R,AB}_{11}(\g;z_1,z_2)
\sim {\delta^{AB} \over \rho^2}}
where $\rho$ is the transverse distance defined in \four. Inserting
the expressions for $\cP^A$ we get a contribution for the interaction
energy of the form
\eqn\snine{ E \sim {1\over \rho^2} \int dt_0~ (\partial_t F_{1~\mu\nu})
(\partial_t F_2^{\mu\nu})}
This is precisely a term of weight four in the effective action of
$SU(2)$ gauge theory, as may
be seen in \rocek. This term is not renormalized since
it is of weight four \dine. Note that this term has two less powers of
$\rho$ in the denominator compared to the $F^4$ terms which
appear in the zero momentum potential. This reflects the
IR-UV connection : an expansion in the time interval gets
translated into an expansion in the transverse distance, which
is the magnitude of the Higgs and hence 
a scale in the effective Yang-Mills theory. Also note
that this is the {\it only} term of weight four other than the $F^4$
terms when fields other than the gauge field are set to zero. 

Terms with higher weight will come from the higher terms of the
Taylor series, from the $F^3$ terms in the 2-form coupling, and
from exchange of other supergravity modes. It would be interesting
to see whether the corresponding operators also obey nonrenormalization
theorems. 

When the brane waves depend on the spatial coordinates on the branes
we expect the time derivatives in \snine\ to be converted into
space-time derivatives on the brane.

Finally, we note that the calculation described above
does not really probe the {\it retarded} nature of the propagator -
an advanced propagator would lead to the same result. 
Strictly speaking we have been investigating
consequences of the finite speed of light rather than causality.
However, we expect that at higher orders the difference of
retarded and advanced propagators will play a role.

\subsec{Branes in $\adsf$}
The calculation outined above is a test for our proposal in the
simplest possible setting. A test of the proposal in the context
of the AdS/CFT correspondence requires an analysis which involves
propagators of fields in the $AdS_5 \times S^5$ background with
nonzero brane momentum. These propagators have been obtained in
full generality \ref\mathur{S. Mathur, {\it private communication}}.
We expect that the signature of causal propagation in the
$AdS_5 \times S^5$ background in terms of the higher derivative
operators in the Yang-Mills effective action would hold in this
case as well. For this to work the cancellation which made the
constant field interaction energy in $\adsf$ equal to the flat
space result should continue to work for non-constant fields.
Our results in this direction will appear in
a future publication \dmn.

\newsec{Acknowledgements}

Many of the issues discussed in this paper arose out of discussions
with Samir Mathur. I am grateful to him for sharing his insights and
for collaboration at the early stages of this work.  I would like to
thank D. Kabat, S. Sethi, D. Tong, S. Trivedi and S. Wadia for
discussions. I also thank S.N. Bose National Center for Basic
Sciences, Calcutta and Saha Institute for Nuclear Physics, Calcutta
for hospitality during the completion of the manuscript.

\newsec{Appendix A : The scalar propagator in flat 6d space}

Consider flat six
dimensional (euclidean) space with the metric
\eqn\aone{dr^2 + r^2 \sum_{i=1}^5 f_i(\theta_i) (d\theta_i)^2}
The action of a complex massless scalar field is
\eqn\atwo{S_\phi = \half \int dr~ d\Omega_5~r^5 ~\phi^* [
{1\over r^5}\partial_r (r^5 \partial_r \phi) 
+ \sum_{i=1}^5{1\over h~r^2}\partial_i(h f_i^{-1}\partial_i \phi)]}
The relevant mode decomposition for this action is
\eqn\athree{\phi (r,\theta_i) = \int {d\beta \over 2\pi} \sum_{k\vm}
{1\over r^2}r^{i\beta}~Z_{k,\vm}(\theta_i)~\phi_{\beta,k,\vm}}
where the $S^5$ (scalar) spherical harmonics $Z_{k,\vm}(\theta_i)$ 
satisfy \van, \ref\matalec{S.D. Mathur and A. Matusis,
hep-th/9805064.}
\eqn\nineteen{
{1\over h}\partial_i({1\over f_i}h \partial_i)Z_{k,\vm}(\theta_i)
= - k(k+4)~Z_{km}(\theta_i)}
with integer $k$ and are chosen to be orthornormal with the
measure $[d\Omega_5]$. 
The action then becomes
\eqn\afour{S_\phi = \int {d\beta \over 2\pi} \sum_{k,\vm}
[\beta^2 + (k+2)^2]\phi^*_{\beta,k,\vm}\phi_{\beta,k,\vm}}
Thus the propagator is
\eqn\afive{G(\vr_1,\vr_2) = \int {d\beta \over 2\pi}\sum_{k,\vm} {1 \over (r_1 r_2)^2}
({r_1 \over r_2})^{i\beta}{1\over \beta^2 + (k+2)^2} Z^*_{k,\vm}
(\theta_1) Z_{k,\vm}(\theta_2)}
Integrating over $\beta$ for $r_1 > r_2$ now gives
\eqn\asix{G(\vr_1,\vr_2) = {\pi \over r_1^4} \sum_{k,\vm}
{1\over 2(k+2)} ({r_2 \over r_1})^k ~ Z^*_{k,\vm}(\theta_1) 
Z_{k,\vm}(\theta_2)}
However we know that the position space propagator in six dimensions
is
\eqn\aseven{G(\vr_1,\vr_2) = {1\over 4\pi^3|\vr_1 - \vr_2|^4}}
Comparing \aseven\ and \asix\ we get 
\eqn\twofive{{1\over 2\pi^3 |\vr_1 - \vr_2|^4}
= {1 \over r_1^4} \sum_{km} {1\over k + 2}~ ({r_2 \over
r_1})^k~Z_{km}^*(\theta_1)~Z_{km}(\theta_2)~~~~~ (r_1 > r_2)} 
This equation can be also proved by using explcit properties of the $S^5$
spherical harmonics.

\newsec{Appendix B : Propagator for 2-forms in $\adsf$}

The mode decomposition which diagonalizes the action \sixteen\ is
\eqn\eighteen{\phi^A_n (x, \theta_i) 
= \int_{-\infty}^\infty {d\beta \over 2\pi} \sum_{k,\vm} e^{-i\beta x}
Z_{k,\vm}(\theta_i)~\phi^A_{n ,(\beta,k,\vm)}~~~~~n= 1,2}

With this mode decomposition and a 
partial integration the action $S_B$ may be diagonalized
to yield
\eqn\twenty{S_B = \half \sum_A \int {d\beta \over 2\pi} \sum_{k\vm}
\pmatrix{\phi^{A~*}_1 & \phi^{A~*}_2}
\pmatrix{(\beta^2 + k(k+4)) & -4\beta \cr
-4\beta & -(\beta^2 + k(k+4))}
\pmatrix{\phi^A_1 \cr \phi^A_2}}
In \twenty\ $\phi^A$ stands for $\phi^A_{(\beta,k,\vm)}$.

The eigenvalues of the kinetic
energy matrix may be easily seen to be
\eqn\twoone{\lambda_\pm = \pm {\sqrt{(\beta^2 + k^2)(\beta^2
+ (k+4)^2)}}}
This clearly shows the two branches of this field found in 
\van. These have masses $k$ and $(k+4)$ respectively.

The propagator may be now found by inverting the matrix. The
result is
\eqn\twotwo{ N^{AB}(\beta k) = {\delta^{AB}
\over (\beta^2 + k^2)(\beta^2 + (k+4)^2)}
\pmatrix{\beta^2 + k(k+4) & -4\beta \cr
-4\beta & -(\beta^2 + k(k+4))}}
The nature of the propagator may be made more transparent by
rewriting the matrix elements of $N$ as
\eqn\twothree{\eqalign{& N^{AB}_{11} (\beta,k) = - N^{AB}_{22}(\beta,k)
= {\delta^{AB}\over 2k + 4}[{k \over \beta^2 + k^2}
+ {k+4 \over \beta^2 + (k+4)^2}] \cr
& N^{AB}_{12}(\beta,k) = N^{AB}_{21}(\beta,k) 
= -{\delta^{AB} \over 2k + 4}[{\beta \over \beta^2 + k^2}
- {\beta\over \beta^2 + (k+4)^2}]}}

The position space propagators may be now easily calculated
\eqn\twothree{N^{AB}_{mn}(\vr_1, \vr_2)
= \delta^{AB}\int {d\beta \over 2\pi} \sum_{km} e^{i\beta(x_1-x_2)}
~Z_{km}^*(\theta_1)
~Z_{km}(\theta_2)~N^{AB}_{mn} (\beta,k)}
The integral over $\beta$ may be now performed to get, for $x_1 > x_2$
\eqn\twofour{\eqalign{&N^{AB}_{11}(\vr_1,\vr_2)
=\delta^{AB}  \sum_{km}{1\over 4(k+2)}
[({r_2 \over r_1})^k + ({r_2 \over r_1})^{k+4}]~
Z_{km}^*(\theta_1)~Z_{km}(\theta_2)\cr
&N^{AB}_{12}(\vr_1,\vr_2)
=-i \delta^{AB} \pi \sum_{km}{1\over 4(k+2)}
[({r_2 \over r_1})^k - ({r_2 \over r_1})^{k+4}]~
Z_{km}^*(\theta_1)~Z_{km}(\theta_2)}}
where we have used \fifteen.
Using the relation \twofive\ in Appendix A, we get
the final answer for the propagator matrix
\eqn\twosix{N^{AB} (\vr_1,\vr_2) = {\delta^{AB}\over 8\pi^3
|\vr_1 - \vr_2|^4}
\pmatrix{(r_1^4 + r_2^4) & -i(r_1^4 - r_2^4) \cr
-i(r_1^4 - r_2^4) & -(r_1^4 + r_2^4)}}
Restoring powers of $R$ yields \twosixa.

\listrefs
\end